\documentclass[nofootinbib,reprint]{revtex4-1} 

\usepackage{graphicx}  
\usepackage{dcolumn}   
\usepackage{bm}        
\usepackage{amssymb}   
\usepackage{color}
\usepackage{setspace}
\usepackage{amsmath}
\usepackage{ulem}
\usepackage{epstopdf}
\usepackage{notes2bib}
\usepackage{caption}
\usepackage{subcaption}

\newcommand\numberthis{\addtocounter{equation}{1}\tag{\theequation}}

\begin{document}

\title{Information theory demonstration of the Richardson cascade}

\author{R.T. Cerbus}
\email{rory.cerbus@oist.jp}
\affiliation{Fluid Mechanics Unit, Okinawa Institute of Science and Technology Graduate University, Okinawa 904-0495, Japan}
\author{W.I. Goldburg}
\affiliation{Department of Physics and Astronomy, University of Pittsburgh, 3941 O'Hara Street, Pittsburgh PA 15260}

\begin{abstract}
Turbulence theory is usually concerned with the statistical moments of the velocity and its fluctuations. One could also analyze the implicit probability distributions. This is the purview of information theory. Here we use information theory, specifically the conditional entropy, to analyze (quasi-)2D turbulence. We recast Richardson's ``eddy hypothesis" that large eddies break up into small eddies in time using the language of information theory. In addition to confirming Richardson's idea, we find that self-similarity and turbulent length scales reappear naturally. Not surprisingly, we also find that the direction of information transfer is the same as the direction of the cascade itself. Consequently, intermittency may be considered a necessary companion to all turbulent flows.
\end{abstract}

\maketitle


\section{Introduction}


The unpredictability of turbulence makes a deterministic analysis of the instantaneous velocity field not only impractical, but very nearly impossible. Researchers have instead studied the statistical properties of turbulence, which necessarily involves the probabilities of velocities and their fluctuations  \cite{tritton1988,tennekes1972}. Although information theory is the natural language for treating these probability distributions \cite{shannon1948,shannon1964,cover1991,brillouin1962}, there have been few studies that make use of it. Instead, the focus has often been on the moments \cite{tennekes1972,frisch1995,pope2000}.

In wall-bounded flows, for example, considerable effort has been directed towards determining the mean velocity profile as a universal function of distance from the wall \cite{tennekes1972,pope2000}. In other situations, the fluctuations are of primary interest and the focus has been on the moments of velocity differences \cite{tennekes1972,frisch1995,pope2000}:
\begin{equation}
S_n(r) = \langle \delta u(r) ^n \rangle = \langle {\big [ (} \vec{u}(\vec{x}+\vec{r}) - \vec{u}(\vec{x}) {\big )} \cdot \hat{r} ] ^n  \rangle_x.
\end{equation}
These velocity differences are thought to represent the characteristic velocity of a turbulent ``eddy" of size $r$, a concept that goes back to L.F. Richardson and his contemporaries \cite{richardson1926,richardson1920}. The importance of $\delta u(r)$ in turbulence is apparent from its appearance in Kolmogorov's $\frac{4}{5}$th law for $S_3(r)$ \cite{kolmogorov1941}. This law is derived using several significant assumptions and the Navier-Stoke's equations \cite{kolmogorov1941,frisch1995} and remains one of the only exact solutions in turbulence. It is the starting point for the entire scaling phenomenology of turbulence.

Here we propose a different approach. Instead of beginning with the moments $S_n(r)$ and Kolmogorov-type assumptions, we will focus on the probability distributions used to calculate the $S_n(r)$, $p(\delta u(r))$ \cite{note1}:
\begin{equation}
S_n(r) = \langle \delta u(r) ^n \rangle = \int p(\delta u(r)) \delta u(r) ^n d(\delta u(r))
\end{equation}
With these probability distributions, information theory can be used to make quantitative statements about, $e.g.$, the unpredictability of $\delta u(r)$. Of course, all the information about the moments is contained in $p(\delta u(r))$, so our analysis can not be completely unrelated to the traditional theory. In order to gauge the usefulness of this approach, we apply it to a key turbulence concept.

Retreat back to the time before Kolmogorov. Consider the ``eddy hypothesis" attributed to Richardson which he summarized in his now famous poem \cite{richardson2007}:

{\centering
\noindent
Big whirls have little whirls \\
That feed on their velocity, \\
And little whirls have lesser whirls \\
And so on to viscosity. \\
}
The cascade concept captured by this rhyme can be formulated with information theory, because the $direction$ of the cascade leaves its mark on the probability distributions of the eddies (whirls). In short, the uncertainty in observing a big eddy become a small eddy should be less than the reverse.

Using experimental observations of (quasi-)2D turbulent flow, we confirm the eddy hypothesis without the Navier-Stoke's equations, scaling arguments or Kolmogorov's assumptions, although some of the features of that kind of analysis (self-similarity \cite{barenblatt2003}) reappear naturally. A completely new result is the existence and direction of information transfer. Before we can come to these conclusions, however, we must review the salient features of turbulence and information theory.

Two-dimensional (2D) turbulence occurs approximately in nearly all large-scale atmospheric flows due partly to the fact that the thickness of the earth's atmosphere is very small compared to its breadth \cite{kellay2002}. This, along with stratification \cite{vallis2006}, result in vast regions of the atmosphere where the vertical velocity is negligible compared with the horizontal. For the same reason large scale oceanic flows are also considered two-dimensional. Our measurements are made using a soap film, which has an even smaller aspect ratio. The physics of soap films and their usefulness in studying 2D flows and 2D turbulence has already been well documented \cite{chakraborty2011,kellay2002,boffetta2012,kraichnan1967,kraichnan1980}, but we present some of the essential experimental details below.  Although we utilize 2D turbulence, our results should extend to 3D turbulence without any significant alteration.

\begin{figure}[h!]
\hspace{-1.5em}
\center
\includegraphics[scale = 0.25]{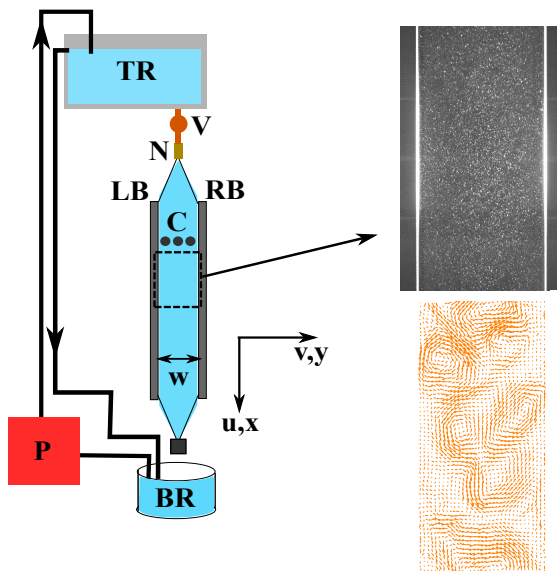}
\caption{Left: Experimental setup showing the reservoirs ($TR$, $BR$), pump ($P$), valve ($V$), nozzle ($N$), comb ($C$), and blades ($LB$, $RB$). Right: Typical particle image used for the PIV analysis with the resulting instantaneous vector field $\vec{u}-\overline{\vec{u}}$ below.}
\label{setup}
\end{figure}

The soap solution is a mixture of Dawn (2$\%$) detergent soap and water with 10 $\mu$m hollow glass spheres added for the velocity measurements. Figure \ref{setup} is a diagram of the experimental setup. The soap film is suspended between two vertical blades connected to a nozzle above and a weight below by nylon fishing wire. The nozzle is connected by tubes to a valve and a reservoir which is constantly replenished by a pump that brings the spent soap solution back up to the reservoir. There is an overflow that leads back to the bottom reservoir so that the height of the top reservoir, and thus the pressure head, is constant. Sometimes the pump feeds directly into nozzle, giving results indistinguishable from the reservoir setup. The flow is always gravity-driven. Typical centerline speeds $\overline{u}$ are several hundred cm/s with rms fluctuations $u'$ generally on the order of 10 cm/s. The channel width $w$ is usually several cm.

The flow velocity is measured using particle image velocimetry (PIV) \cite{adrian2011}. A bright white light source is placed behind the soap film and shines on the soap film at an angle (so that the light does not directly enter the camera, which is perpendicular). The particles scatter more light than the surrounding soap water and are easily distinguished. A fast camera tracks their movement and standard PIV techniques are used to calculate the velocity field. Because the full velocity field at different times is needed for our analysis, we use PIV and can not use, $e.g.$, laser Doppler velocimetry.

We focus on the wall-normal or horizontal velocity $v$ instead of the vertical velocity $u$ because it does not suffer from aliasing effects \cite{tennekes1972} and because $\overline{u}$ changes slightly with $x$ (due to gravitational acceleration). We take velocities very near the center of the channel to avoid wall effects and variations in the energy and enstrophy (norm of vorticity) injection rates \cite{pope2000}.

Turbulence in the soap film is generated by inserting a row of rods (comb) perpendicular to the film. When this protocol is used we almost always observe the direct enstrophy cascade \cite{kellay2002,boffetta2012}. In the traditional phenomenology, enstrophy is transferred from some large injection scale $L_0$ $downscale$ to a viscous dissipative scale $\eta$ \cite{boffetta2012,kellay2002,kraichnan1980}. The rate of enstrophy transfer is $\beta$, which should be constant over the inertial range of scales in between $L_0$ and $\eta$. We introduce this traditional framework in order to contrast it with our own approach, as well as provide evidence to the portential skeptic that we are indeed working with a direct cascade.

\begin{figure}[h!]
\centering
\includegraphics[scale = 0.33]{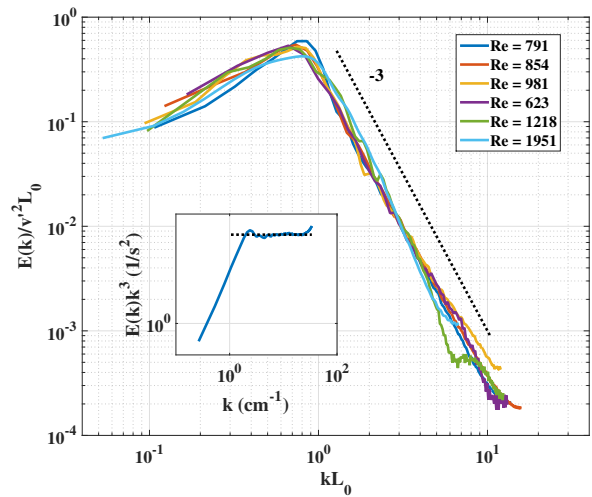}
\caption{The spectra $E(k)$ from the soap film experiments. The curves are normalized using the large length and velocity scales $L_0$ and $v'$ so that for all $Re$ they collapse at low to intermediate $k$ \cite{pope2000}. Inset: A typical plot of $E(k)k^{3}$ which is $\propto \beta^{2/3}$ in the inertial range. This is used to estimate $\beta$.}
\label{spectra}
\end{figure}

As is customary we assume that the energy spectrum $E(k)$ of velocity fluctuations scales with $\beta$ and the wavenumber $k$, an inverse length, for a certain range of $k$ (inertial range). For the direct enstrophy cascade, $E(k) \propto \beta^{2/3} k^{-3}$. In Fig. \ref{spectra} we plot all of the $E(k)$ calculated from $v$. The curves have been normalized using the large length scale $L_0 = \int_0^{\infty} E(k) dk {\big /} \int_0^{\infty} k E(k) dk$ and the rms velocity $v' = \sqrt{2\int_0^{\infty} E(k) dk}$. All of the curves collapse at low and intermediate $k$, a signature of the cascade's self-similarity \cite{barenblatt2003}. The physical process that created this curve is similar in all cases.
Here $Re$ is calculated as $Re = v'L_0/\nu$, where $\nu$ is the kinematic viscosity of water. We will see that a kind of self-similarity also arises naturally in the information theory analysis.

\begin{figure}[h!]
\hspace{-1.5em}
\center
\includegraphics[scale = 0.4]{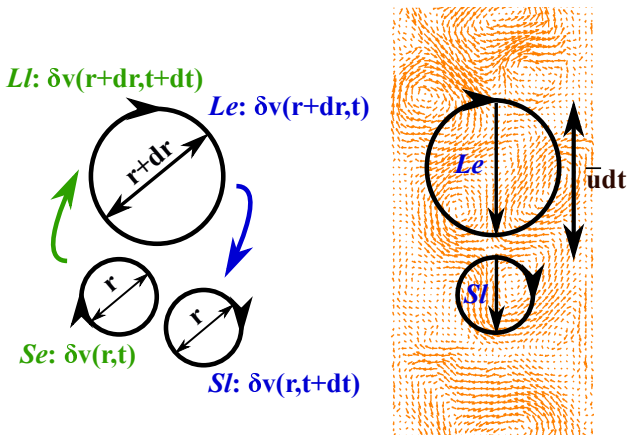}
\caption{A diagram of the four different eddies we consider. The time $t$ is simply a reference. The $dt$ is the time between the large and small eddies considered, and $dr$ is the difference in size between them. We chose $dt$ and $dr$ to maximize the conditional entropies, but the shape and sign of these curves is unaffected. We make our comparison in the frame of reference with mean velocity $\overline{u}$, making this a quasi-Lagrangian measurement \cite{biferale2011}.}
\label{cascade}
\end{figure}

We now outline our test. Consider the diagram of a turbulent cascade shown in Fig. \ref{cascade}, which is typical of that shown in many turbulence textbooks \cite{frisch1995}. The Richardson picture involves eddies evolving in both space and time. A large eddy at an early time ($Le$) will become a small eddy at a later time ($Sl$). (We also consider the reverse process.) We should be more certain about $Sl$ given $Le$ than $Ll$ given $Se$. Let us now move on to the precise description.

The Shannon entropy is central to information theory \cite{cover1991}. We could simply look at the raw probability distributions themselves, but the entropy gives a single number that characterizes how random the distribution is and provides an interpretation in terms of information. For the eddy $Se = \delta v(r,t)$, the Shannon entropy is
\begin{equation}
H(Se) = -\sum_{Se} p(Se) \log_2 p(Se)
\end{equation}
where the sum is over all possible values of $Se$. This is the amount of information we gain from or uncertainty we had prior to measuring $Se$ \cite{cover1991}. (Uncertainty and information are the same in this framework.)

To test a relationship between two eddies, we use a modified form of the entropy: the conditional entropy \cite{cover1991}. This gives us the uncertainty of one eddy given that the other eddy occurred. For the uncertainty of $Sl$ given $Le$ we write
\begin{equation}
H(Sl|Le) = -\sum_{Sl,Le} p(Sl,Le) \log_2 p(Sl|Le).
\label{eq:condH}
\end{equation}
where $p(Sl,Le)$ and $p(Sl|Le)$ are the joint and conditional probabilities respectively. If $Le$ and $Sl$ are independent, $H(Sl|Le) = H(Sl)$. Knowing $Le$ doesn't help us reduce our uncertainty about $Sl$. If $Sl$ is determined by $Le$ with absolute certainty, then $H(Sl|Le) = 0$. So the stronger the relationship, the smaller this quantity is. 

The $\delta v(r)$ take on a continuous range of values, but to calculate probabilities and then estimate entropies, we must bin (discretize) the measured data. This is, in fact, unavoidable due to the finite resolution of all measurement apparatuses. We systematically varied the bin size, but as in Ref. \cite{cerbus2013}, we found our results to be extremely robust. None of our results will change except for a vertical shift in all the conditional entropies by the same factor. We use a bin size of $\sqrt{\overline{\delta v(r)^2}}/3$, which of course changes size with $r$, but ensures that the number of bins (and thus the maximum possible value of the conditional entropy) remains roughly the same.


We assert that $H(Le | Sl) > H(Sl | Le)$ signifies a cascade from large to small scales, and we use $D(r) \equiv H(Le | Sl) - H(Sl | Le) > 0$ to test this. $D(r)$ in principle also depends on $t$, $dt$ and $dr$ as indicated in Fig. \ref{cascade}, but we maximize with respect to $dt$ and $dr$ (experimental parameters) and take many realizations at different $t$, which is simply a reference marker, so that in the end $D(r)$ only depends on $r$. This means that we are formulating Richardson's hypothesis $locally$ at a scale $r$. We note that our approach is similar to work on information transport in spatiotemporal systems \cite{vastano1988}. A study that more closely anticipates our own is an information transfer treatment of the GOY model \cite{materassi2014}.

\begin{figure*}
    \centering
    \begin{subfigure}[t]{0.48\textwidth}
        \centering
        \includegraphics[scale = 0.30]{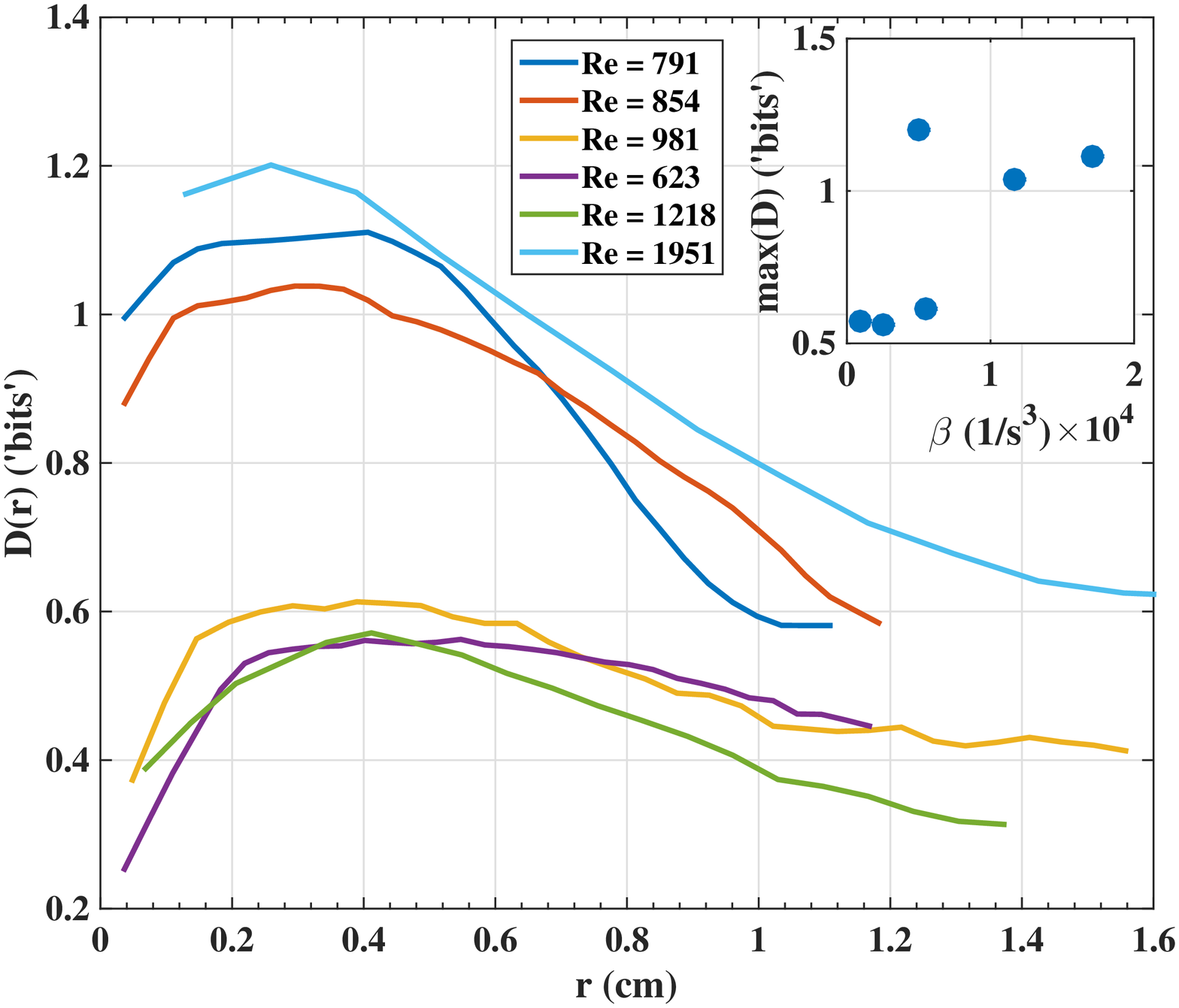}
        \caption{The quantity $D (r)$ plotted vs. $r$ for several different values of $Re$. The positivity of $D (r)$ means that big eddies are on the average becoming small eddies.}
         \label{cond_entropy_a}
    \end{subfigure}
    \begin{subfigure}[t]{0.48\textwidth}
            \centering
        \includegraphics[scale = 0.30]{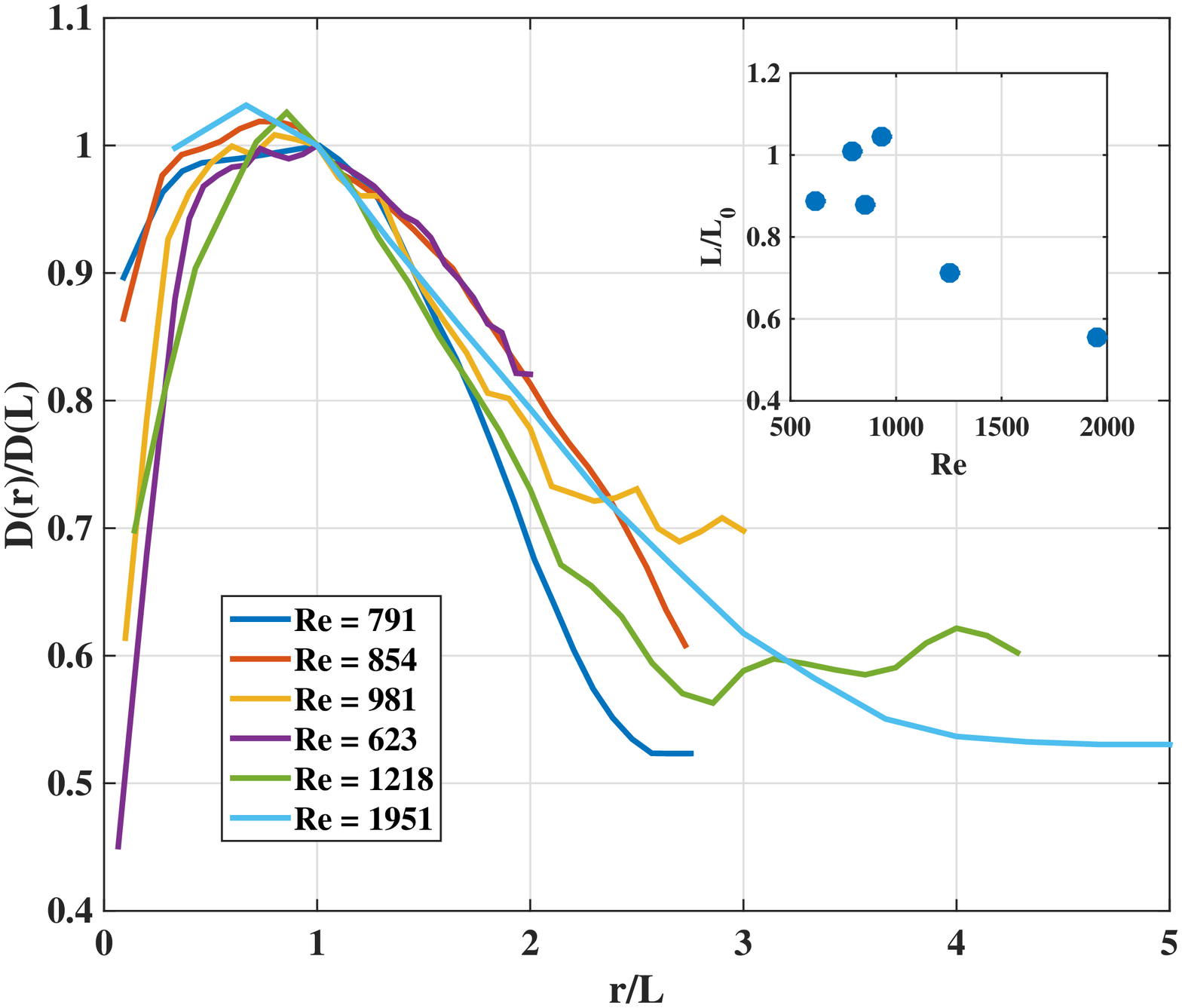}
        \caption{The same curves as in (a), but now normalized using a single length scale $L$. The curves now collapse quite well, although not at large $r/L$.}
        \label{cond_entropy_b}
    \end{subfigure}
    \caption{The information theory representation of Richardson's eddy hypothesis.}
\end{figure*}

The result of the calculation of $D(r)$ is shown in Fig. \ref{cond_entropy_a} for the same $Re$ and $E(k)$ as in Fig. \ref{spectra}. Clearly $D (r)$ is greater than zero in all cases as expected for a direct cascade. In the traditional framework, a cascade with constant transfer rate is first assumed and then the scalings are derived with the extra assumptions (universality, etc.) already mentioned \cite{frisch1995}. Then if the moments scale as predicted, the cascade is considered to be demonstrated. Here we bypass this sophisticated argumentation and show the cascade straightaway.


%
%

More can be gleaned from Fig. \ref{cond_entropy_a}. There appears to be a region of nearly constant value in many of the curves, which is reminiscent of the enstrophy (energy) flux which takes on a maximum and constant value in the inertial range equal to the injection rate $\beta$ ($\epsilon$) \cite{pope2000}. This suggests a connection between max($D$) and $\beta$. Estimating $\beta$ from Fig. \ref{spectra}, we find a general increase of max($D$) with $\beta$.

Moreover, all of the curves appear to have a similar shape. This is reminiscent of the energy spectra above, and suggestive of an underlying self-similarity. Indeed by choosing a single $L$ and normalizing $r$ by $L$ and $D(r)$ by $D(L)$, we find reasonable collapse as seen in Fig. \ref{cond_entropy_b}. Not only have we rediscovered self-similarity, but also a turbulent length scale. $L$ is very close to $L_0$, but their ratio decreases with $Re$ as shown in the inset of Fig. \ref{cond_entropy_b}.

Now let us establish an interesting corollary. The mutual information is the information shared between two variables. For the large eddies at earlier time ($Le$) and the small at later time ($Sl$), we write
\begin{equation}
I(Sl;Le) = H(Sl) - H(Sl | Le) = I(Le;Sl)
\end{equation}
If we also define a mutual information for $Se$ and $Ll$, then we can rewrite $D$ as
\begin{align*}
D = H(Ll | Se) - H(Sl | Le) = H(Ll) - H(Sl) - \\ 
I(Ll;Se) + I(Sl;Le) \approx I(Sl;Le) - I(Ll;Se) \numberthis
\end{align*}
where we have used that $H(Sl) \approx H(Ll)$, due to the small $dr$ (experimentally verified). This means that 
\begin{equation}
I(Sl;Le) - I(Ll;Se) \approx D > 0.
\label{eq:mutinfo}
\end{equation}
In other words, the information shared between eddies going downscale is more than the reverse. There is net information being transferred downscale, concurrently with the enstrophy. This result should apply generally to both the 2D and 3D direct cascades and even the 2D inverse energy cascade. It rests only on the validity of our information theory expression of Richardson's eddy hypothesis, which we have here experimentally verified.

The implications of this result are quite powerful. Kolmogorov's small scale universality assumption can be expressed as the small scales ``forgetting" about the large (forcing) scales \cite{frisch1995}. Eq. \ref{eq:mutinfo} suggests that this can never be true. So long as there is a cascade, there must be information transfer in the same direction which makes ``forgetting" impossible. Thus intermittency appears to be a necessary feature of all turbulent cascades.

We have shown that we can formulate and test Richardson's idea of a cascade without using the Navier-Stoke's equation, scaling arguments or Kolmogorov's assumptions. And yet some of the old patterns, such as self-similarity, re-emerged. This work presents an entirely new perspective on the statistics of turbulent velocities and suggests a more suitable framework for understanding intermittency. Our application of information theory to turbulence ought to serve as a guidepost for further work.

We thank Michael Kister and Chien-chia Liu for many useful discussions. R.T.C. acknowledges support from the Okinawa Institute of Science and Technology Graduate School.



\begin{thebibliography}{99}
\bibitem{note1} Of course with experimental or DNS data one typically bypasses calculating $p(\delta u(r))$ by assuming all samples are taken from this distribution.
\end{thebibliography}
\end{document}